\documentstyle[12pt]{article}

\def\ar{\rightarrow}
\def\bib{\bibitem}

\def\dsl{\partial\!\!\!/}
\def\Dsl{D\!\!\!\!/}
\def\intx{\int\! d^{\sl 4}x}
\def\inty{\int\! d^{\sl 4}y}
\def\intX{\int\! d^{\sl 4}X\,}
\def\intY{\int\! d^{\sl 4}Y\,}

\def\intk{\int\! \frac{d^{\sl 4}k}{(2{\pi})^4}}
\def\intK{\int\! \frac{d^{\sl 4}K}{(2{\pi})^4}\,}

\def\ksl{k\!\!\!/}

\def\Lp{L_{_P}}
\def\midd{\! \mid \!}

\def\pa{\partial}

\def\rvec{\!\!\!\!^{^\rightarrow}}
\def\lvec{\!\!\!\!^{^\leftarrow}}

\def\tr{\,\mbox{tr}\,}

\def\Vreg{V\!\!\!\!\!^{_{reg}}}

\def\al{\alpha}
\def\be{\beta}
\def\ga{\gamma}
\def\de{\delta}
\def\ep{\varepsilon}

\def\th{\vartheta}

\def\la{\lambda}

\def\si{\sigma}

\def\Ga{{\it\Gamma}}
\def\La{{\it\Lambda}}
\def\Om{{\it\Omega}}

\def\PSI{{\it\Psi}}

\def\beq{\begin{equation}}
\def\eeq{\end{equation}}
\def\bed{\begin{displaymath}}
\def\eed{\end{displaymath}}
\def\beqq{\begin{eqnarray}}
\def\eeqq{\end{eqnarray}}
\def\bedd{\begin{eqnarray*}}
\def\eedd{\end{eqnarray*}}

\begin{document}

\centerline{\normalsize\bf SCATTERING CROSS-SECTIONS IN QUANTUM GRAVITY}
\centerline{\normalsize\bf - THE CASE OF MATTER-MATTER SCATTERING}

\vspace*{0.9cm}
\centerline{\footnotesize C. WIESENDANGER}
\baselineskip=12pt
\centerline{\footnotesize\it Aurorastr. 24, CH-8032 Zurich}
\centerline{\footnotesize E-mail: christian.wiesendanger@ubs.com}

\vspace*{0.9cm}
\baselineskip=13pt
\abstract{Viewing gravitational energy-momentum $p_G^\mu$ as equal by observation, but different in essence from inertial energy-momentum $p_I^\mu$ naturally leads to the gauge theory of volume-preserving diffeormorphisms of a four-dimensional inner space. To analyse scattering in this theory the gauge field is coupled to two Dirac fields with different masses. Based on a generalized LSZ reduction formula the $S$-matrix element for scattering of two Dirac particles in the gravitational limit and the corresponding scattering cross-section are calculated to leading order in perturbation theory. Taking the non-relativistic limit for one of the initial particles in the rest frame of the other the Rutherford-like cross-section of a non-relativistic particle scattering off an infinitely heavy scatterer calculated quantum mechanically in Newtonian gravity is recovered. This provides a non-trivial test of the gauge field theory of volume-preserving diffeomorphisms as a quantum theory of gravity.}

\normalsize\baselineskip=15pt

\section{Introduction}

Imagine a world in which physicists would be forced from the outset to think about gravitation in terms and in the language of relativistic quantum field theory - a language consisting of terms such as state vectors in Fock spaces, causal quantum fields, operators, probability amplitudes, observables, propagators, conserved quantities such as the electric charge, energy-momentum and the like. Within that framework they might try to answer questions such as "Given a certain number of incoming particles described by free state vectors with given inertial energy-momenta $ p_i^\mu $ and other quantum numbers what is the probability - after they have interacted gravitationally - to observe a certain number of outgoing particles described by free state vectors with measured inertial energy-momenta $ p_f^\mu $ and other quantum numbers?" - or to construct the $S$-matrix for quantum gravity.

To answer such questions imagine those physicists following a similar reasoning as in quantum electrodynamics to construct a theory yielding the $S$-matrix for quantum gravity. Hence they would start with asymptotic states and fields describing matter as in the case of QED implementing microcausality and space-time symmetries from the outset. And as in the case of QED they would look for a conserved quantity related to a global gauge symmetry which could generate the gravitational interaction through gauging the symmetry locally - in the case of QED it is the electric charge related to a global $U(1)$-symmetry the gauging of which yields the gauge field $ A_\mu $ transmitting the electromagnetic interaction. In addition they would note that in their approach spacetime with its Minkowski geometry is a background A Priori and becomes "visible" only indirectly e.g. via the validity of relativistic relations such as $ p^2 = m_I^2 $ for an observable particle of inertial mass $ m_I $.

To identify a conserved quantity which our physicists could relate to a gauge field transmitting the gravitational interaction at the quantum field level they would have to go back to the very outset of what is known about gravity. Ultimately this is the observed equality of inertial and gravitational mass $ m_I = m_G $. To be in agreement with observation this equality has to hold in any expression describing observable states in a gravitational context in their rest frames. However, our physicists could argue that nothing enforces this equality to hold for virtual (=non-observable) quantum states as long as it continues to hold for the on-shell (=observable) quantum states.

Now (a) the observed equality of inertial and gravitational mass of an on-shell physical object in its rest frame together with (b) the conservation of the inertial energy-momentum $ p_I^\mu $ for an asymptotic state in any reference frame tells our physicists that in the rest frame
\beq \label{0}
p_I^\mu = (m_I,\underline{0}) =^{\!\!\!\!\!\!\! ^{ (a)}} (m_G,\underline{0}) = p_G^\mu
\eeq
assuming that the gravitational energy-momentum $ p_G^\mu $ plays a physical role different from that of the inertial energy-momentum, yet being observationally identical for on-shell objects. In addition they might argue that there could in fact be two separate conservation laws for off-shell states, one for the inertial energy-momentum and the other for the gravitational energy-momentum.

To explore this route our physicists might postulate that both $p_I^\mu$ and $p_G^\mu$ are two separate four-vectors which are conserved for any asymptotic state, but in their approach through two different mechanisms. The conservation of $p_I^\mu$ would be related to translation invariance in spacetime as usual. Making use of Noether's theorem a second conserved four-vector could then be constructed which is related to the invariance under volume-preserving diffeomorphisms of a four-dimensional inner space. That four-vector would then be interpreted as the gravitational energy-momentum $p_G^\mu$ in the construction of a gauge theory of gravitation.

Our physicists would finally assure the observed equality of inertial and gravitational energy-momentum for on-shell observable physical objects in this approach by taking the gravitational limit, i.e. equating both types of momenta. They would also note - being forced from the outset to think about gravitation in terms and in the language of relativistic quantum field theory - that the language of classical physics with its reference to spacetime trajectories of particles makes no sense in the context of constructing such a theory - as would the principle of equivalence. Only in the limit of $ \hbar \ar 0 $ should both re-emerge.

In fact, we have worked out the above line of thinking on the basis of which we have defined the classical and quantum gauge field theories of the group of volume-preserving diffeomorphisms \cite{chw1,chw2} and proven its renormalizability \cite{chw3}. Separately we have specified the asymptotic observable states of the theory, its $S$-matrix and its LSZ reduction formulae - all taking into account the aforementioned gravitational limit \cite{chw4}. Finally we have analyzed the classical limit $ \hbar \ar 0 $ in which the original gauge symmetry under volume-preserving diffeomorphisms of inner space disappears and invariance under general spacetime coordinate transformations emerges - and with it general relativity \cite{chw5}. This is the basis of our claim that the gauge theory of volume-preserving diffeomorphisms of an inner Minkowski space is a viable, renormalizable theory of quantum gravity.

This being the case we can now directly analyze physical situations for which we can compare predictions both within the framework of the theory presented as well as within the standard framework of Newtonian gravity dealt with quantum-mechanically such as the gravitational scattering of two particles with different masses. 

Hence in this paper we calculate the scattering cross-section of two Dirac particles with different masses and compare it in an appropriate limit with the cross-section of a non-relativistic particle scattering off an infinitely heavy scatterer calculated quantum mechanically in Newtonian gravity - and determine the numerical value of the coupling constant of our theory in the process.

\section{Matter-Matter Scattering Amplitude}
In this section we calculate the scattering amplitude of two Dirac particles with different masses in quantum gravity to lowest order in perturbation theory in natural units $c = \hbar = 1$.

Our starting point is the action for two Dirac fields $\psi$ and $\PSI$ with masses $m$ and $M$ respectively which are coupled to the gravitational field $A^\mu\,_\al$ as we have generally defined it in \cite{chw1,chw2,chw4}
\beqq \label{1}
S &=& \intx \intX\La^{-4}\, \Bigg\{
\frac{1}{4}\, F_{\mu\nu}\,^\al(x,X) \cdot F^{\mu\nu}\,_\al(x,X) \nonumber \\
& &\quad +\, \frac{\la}{2}\, \pa_\mu A^\mu\,_\al(x,X)
\cdot \pa^\nu A_\nu\,^\al(x,X) \nonumber \\
& &\quad -\, \frac{\mu^2}{2}\, A^\mu\,_\al(x,X) \cdot A_\mu\,^\al(x,X) \\
& &\quad +\, {\overline\psi}(x,X) \, \left(\frac{i}{2}\Dsl\,\rvec
- \frac{i}{2}\Dsl\,\lvec - m \right) \psi(x,X) \nonumber \\
& &\quad +\, {\overline\PSI}(x,X) \, \left(\frac{i}{2}\Dsl\,\rvec
- \frac{i}{2}\Dsl\,\lvec - M \right) \PSI(x,X) \Bigg\}. \nonumber
\eeqq
Above, $x$ and $X$ denote spacetime and inner space coordinates \cite{chw1,chw2,chw4},
\beqq \label{2}
& &\quad\quad\quad\quad F_{\mu\nu}\,^\al(x,X) = \pa_\mu A_\nu\,^\al(x,X) - \pa_\nu A_\mu\,^\al(x,X) \\
& & +\, g \La\, A_\mu\,^\be(x,X) \cdot \nabla_\be A_\nu\,^\al(x,X) - g \La\, A_\nu\,^\be(x,X) \cdot \nabla_\be A_\mu\,^\al(x,X) \nonumber
\eeqq
denotes the gravitational field strength and
\beq \label{3}
D_\mu = \pa_\mu + g \La\, A_\mu\,^\al \cdot \nabla_\al \eeq
the covariant derivative \cite{chw1,chw2,chw4}, $g$ a dimensionless coupling constant and $\La$ a length scale in inner space \cite{chw1}. Note that we have written down the action in a so-called Minkowski gauge and have added both a gauge-fixing term for the remaining gauge degrees of freedom proportional to a constant $\la$ and a mass term for the gauge field with mass $\mu$ to deal with possible infrared problems \cite{chw4}. All other notations and conventions have been collected in Appendix A.

We want to calculate the scattering amplitude of two Dirac particles with incoming and outgoing inertial equal to gravitational energy-momenta $p_i, q_i$, and $p_f, q_f$ respectively, incoming and outgoing spins $\ga_i, \ga_i'$ and $\ga_f, \ga_f'$ respectively and masses $m$ ($p_i^2 = p_f^2 = m^2$) and $M$ ($q_i^2 = q_f^2 = M^2$). Above $i$ and $f$ refer to initial and final states.

In quantum gravity $S$-matrix elements are related by generalized LSZ reduction formulae \cite{chw4} to the gravitational limit of truncated on shell Fourier-transformed vacuum expectation values of time-ordered products of field operators in the interacting theory. Applying the general expression Eqn.(150) for generalized Dirac matter LSZ reduction formulae in \cite{chw4} to the case at hands the amplitude is found to be
\newpage
\beqq \label{4}
& & \langle p_f, q_f\, \mbox{out}\midd p_i, q_i\, \mbox{in} \rangle
= \lim_{\mu \ar 0} \lim_{P_f \ar p_f} \lim_{P_i \ar p_i}
\lim_{Q_f \ar q_f} \lim_{Q_i \ar q_i}
\nonumber \\
& &\quad \left( \frac{i}{\sqrt Z_2} \right)^2
\intx_i \intX\!_i\, \La^{-4}\,
\intx_f \intX\!_f\, \La^{-4} \nonumber \\
& &\quad \left( \frac{i}{\sqrt Z_2} \right)^2
\inty_i \intY\!_i\, \La^{-4}\,
\inty_f \intY\!_f\, \La^{-4} \nonumber \\
& &\quad\quad\quad\quad {\overline u} (p_f,\ga_f)\,
e^{ip_f x_f + iP_f X_f} \Big( i{\dsl\rvec}_{x_f} - m \Big) \\
& &\quad\quad\quad\quad {\overline U} (q_f,\ga'_f)\,
e^{iq_f y_f + iQ_f Y_f} \Big( i{\dsl\rvec}_{y_f} - M \Big)
\nonumber \\
& &\quad \langle 0 \midd
T\Big({\overline \psi} (x_i,X_i)\, \psi (x_f,X_f)\,
{\overline \PSI} (y_i,Y_i)\, \PSI (y_f,Y_f) \Big)
\midd 0 \rangle \nonumber \\
& &\quad\quad\quad\quad \Big( -i{\dsl\lvec}_{x_i} - m \Big)
u (p_i,\ga_i)\, e^{-ip_i x_i - iP_i X_i} 
\nonumber \\
& &\quad\quad\quad\quad \Big( -i{\dsl\lvec}_{y_i} - M \Big)
U (q_i,\ga'_i)\, e^{-iq_i y_i - iQ_i Y_i}.
\nonumber
\eeqq
Above $u(p,\ga)$ and $U(q,\ga')$ denote free Dirac spinors describing the asymptotic states of the fields $\psi$ and $\PSI$ with momenta $p, q$ and spins $\ga, \ga'$ respectively and $Z_2$ the spinor field renormalization constant.

Next we have to calculate the time-ordered product of the four interacting field operators in Eqn.(\ref{4}) which is obtained from the generating functional ${\cal Z}\left[\eta, {\overline \eta};H, {\overline H}; J \right]$ for the Green functions in quantum gravity by
\beqq \label{5}
& &\quad\quad \langle 0 \midd
T\Big({\overline \psi} (x_i,X_i)\, \psi (x_f,X_f)\,
{\overline \PSI} (y_i,Y_i)\, \PSI (y_f,Y_f) \Big)
\midd 0 \rangle = \\
& & \frac{\La^4\, \de\,\rvec}{i\, \de {\overline \eta} (x_f,X_f)}
\frac{\La^4\,  \de\,\rvec}{i\, \de {\overline H} (y_f,Y_f)}
{\cal Z}\left[\eta, {\overline \eta};H, {\overline H}; J \right]
\frac{\La^4\,  \de\,\lvec}{i\, \de \eta (x_i,X_i)}
\frac{\La^4\,  \de\,\lvec}{i\, \de H (y_i,Y_i)}, \nonumber
\eeqq
where $\eta, H, J$ denote external sources for the fields $\psi, \PSI, A$ to which they are coupled through linear terms in the action. ${\cal Z}$ has been defined in \cite{chw2} in the path integral representation Eqn.(46) in that paper. We now turn to evaluate it perturbatively in the usual way
\beqq \label{6}
{\cal Z}\left[\eta, {\overline \eta};H, {\overline H}; J \right]
&=& \exp\,i\, S_{INT} \left[
\frac{\La^4\, \de\,\rvec}{i\, \de \eta},
\frac{\La^4\, \de\,\rvec}{i\, \de {\overline \eta}};
\frac{\La^4\, \de\,\rvec}{i\, \de H},
\frac{\La^4\, \de\,\rvec}{i\, \de {\overline H}};
\frac{\La^4\, \de\,\rvec }{i\, \de J}
\right] \\
& &\quad {\cal Z}_{\sl 0} \left[\eta, {\overline \eta};H, {\overline H}; J \right]. \nonumber
\eeqq

Above ${\cal Z}_{\sl 0}$ is the generating functional for free Green functions as given by Eqn.(50) in \cite{chw2} and $S_{INT}$ the interaction part of the action $S$ in Eqn.(\ref{1}) cubic and quartic in the fields. ${\cal Z}_{\sl 0}$ is easily calculated to be \cite{chw2}
\beqq \label{7}
& & {\cal Z}_{\sl 0} \left[\eta, {\overline \eta};H, {\overline H}; J \right]
\propto
\exp \frac{i}{2} \int\!\!\int J_\mu\,^\al \cdot
G_F^{\mu\nu}\,_{\al\be} \, J_\nu\,^\be \\
& &\quad\quad \exp\,i \int\!\!\int {\overline \eta} \cdot S_F^m \, \eta
\cdot \exp\,i \int\!\!\int {\overline H} \cdot S_F^M \, H,
\nonumber
\eeqq
where $S_F^m$ and $G_F^{\mu\nu}\,_{\al\be}$ denote the free Dirac propagator as in Eqn.(66) in \cite{chw4}
\beq \label{8}
S_F^m (x, X) = i\, \La^4\, \de^4(X) \intk\, e^{-ik\, x}\, \frac{\ksl + m}{k^2 - m^2 + i\ep}
\eeq
and gauge field propagator as in Eqn.(103) in \cite{chw4} for the choice of gauge made in that paper
\beq \label{9}
G_F^{\mu\nu}\,_{\al\be} (x, X)
= -i\,\, ^T\!\de_{\al\be}(X)
\intk\, e^{-ik\, x}\, \frac{\eta^{\mu\nu}}{k^2 - \mu^2 + i\ep}
\eeq
for the fields $\psi$ and $A_\mu\,^\al$ respectively with an expression analogous to Eqn.(\ref{8}) for the propagator of the field $\PSI$. Above $^T\!\de_{\al\be}(X)$ refers to the delta function transversal in inner space
\beq \label{10}
^T\!\de_{\al\be}(X) = \intK \La^4\, e^{-iK\, X}
\left( \eta_{\al\be} - \frac{K_\al K_\be}{K^2}\right)
\eeq
introduced in \cite{chw4}.

The part of $S_{INT}$ relevant to our calculation is
\beqq \label{11}
& & S_{INT} = \intx \intX\La^{-4}\, \frac{i g \La}{2}\, \Big\{
\, \mbox{other terms} \nonumber \\
& &\quad +\, {\overline\psi}(x,X) \, 
\left( \ga^\mu A_\mu\,^\al(x,X) {\nabla\rvec}_\al
- \ga^\mu {\nabla\lvec}_\al A_\mu\,^\al(x,X) \right) \psi(x,X) \\
& &\quad +\, {\overline\PSI}(x,X) \, 
\left( \ga^\mu A_\mu\,^\al(x,X) {\nabla\rvec}_\al
- \ga^\mu {\nabla\lvec}_\al A_\mu\,^\al(x,X) \right) \PSI(x,X) \Big\}. \nonumber
\eeqq
Note the arrows on the derivatives w.r.t. inner coordinates indicating the directions in which they act.

Evaluating the functional derivatives in Eqns.(\ref{5}) and (\ref{6}), setting the source terms equal to zero and discarding disconnected and higher order contributions we obtain the vacuum expectation value for the time-ordered product of the four field operators to leading order
\beqq \label{12}
& &\langle 0 \midd
T\Big({\overline \psi} (x_i,X_i)\, \psi (x_f,X_f)\,
{\overline \PSI} (y_i,Y_i)\, \PSI (y_f,Y_f) \Big)
\midd 0 \rangle = \nonumber \\
& &\quad\quad\quad - \left(\frac{i g \La}{2}\right)^2\, 
\intx \intX\!\, \La^{-4}\, \inty \intY\!\, \La^{-4} \nonumber \\
& &\quad \Big[ S_F^m (x_f - x, X_f - X) \ga_\mu
\left({\nabla\rvec}_x^\al S_F^m (x - x_i, X - X_i) \right)
\nonumber \\
& &\quad - \left( S_F^m (x_f - x, X_f - X) \ga_\mu
{\nabla\lvec}_x^\al \right) S_F^m (x - x_i, X - X_i) \Big]
\\
& &\quad\quad\quad\quad\quad\quad
i\, G_F^{\mu\nu}\,_{\al\be} (x - y, X - Y)
\nonumber \\
& &\quad \Big[ S_F^M (y_f - y, Y_f - Y) \ga_\nu
\left({\nabla\rvec}_y^\be S_F^M (y - y_i, Y - Y_i) \right)
\nonumber \\
& &\quad - \left(S_F^M (y_f - y, Y_f - Y) \ga_\nu
{\nabla\lvec}_y^\be \right) S_F^M (y - y_i, Y - Y_i) \Big].
\nonumber
\eeqq

Inserting this expression in Eqn.(\ref{4}) and performing the truncation we find the scattering amplitude to be
\beqq \label{13}
& & \langle p_f, q_f\, \mbox{out}\midd p_i, q_i\, \mbox{in} \rangle
= \lim_{\mu \ar 0} \lim_{P_f \ar p_f} \lim_{P_i \ar p_i}
\lim_{Q_f \ar q_f} \lim_{Q_i \ar q_i}
\, i (g \La)^2\, \nonumber \\
& &\intx \intX\!\, \La^{-4}\, \inty \intY\!\, \La^{-4}
\intk \intK \La^4 \nonumber \\
& &\quad\quad\quad\quad {\overline u} (p_f,\ga_f)\,
\frac{i(P_i^\al + P_f^\al)}{2}\, \ga_\mu\, u (p_i,\ga_i) \\
& &\quad\quad\quad\quad\quad
\frac{\eta^{\mu\nu}}{k^2 - \mu^2 + i\ep}
\left( \eta_{\al\be} - \frac{K_\al K_\be}{K^2}\right)
\nonumber \\
& &\quad\quad\quad\quad {\overline U} (q_f,\ga'_f)\,
\frac{i(Q_i^\be + Q_f^\be)}{2}\, \ga_\nu\, U (q_i,\ga'_i)
\nonumber \\
& &\quad
e^{ix( p_f - p_i - k )}\,
e^{iy( q_f - q_i + k )}\,
e^{iX( P_f - P_i - K )}\,
e^{iY( Q_f - Q_i + K )}.
\nonumber 
\eeqq
Performing the remaining integrations the amplitude finally becomes
\beqq \label{14}
& & \langle p_f, q_f\, \mbox{out}\midd p_i, q_i\, \mbox{in} \rangle
= \lim_{\mu \ar 0} \lim_{P_f \ar p_f} \lim_{P_i \ar p_i}
\lim_{Q_f \ar q_f} \lim_{Q_i \ar q_i}
\, (-i)\, (g \La)^2\, \nonumber \\
& &\quad 
(2\pi)^4\, \de^4(p_f - p_i + q_f - q_i)\,
(2\pi)^4\, \La^{-4}\, \de^4(P_f - P_i + Q_f - Q_i) \quad\quad \\
& & {\overline u} (p_f,\ga_f)\, \ga^\mu\, u (p_i,\ga_i)
\, \frac{1}{(p_f - p_i)^2 -\mu^2 + i\ep}\, 
{\overline U} (q_f,\ga'_f)\, \ga_\mu\, U (q_i,\ga'_i) \nonumber \\
& &\quad\quad \frac{P_{i\al} + P_{f\al}}{2}\, 
\left( \eta^{\al\be} - \frac{ (P_f^\al - P_i^\al)
(Q_i^\be - Q_f^\be) }{ (P_f - P_i) (Q_i - Q_f) }\right)\,
\frac{Q_{i\be} + Q_{f\be}}{2}.
\nonumber
\eeqq
Note that before taking the gravitational limit the amplitude is scale-invariant under $P \ar \rho P$, $Q \ar \rho\, Q$ and $\La \ar \rho^{-1} \La$ as it has to be \cite{chw1,chw2}.

Trying to take the limits above we are left with an expression of the type $(2\pi)^4\, \La^{-4}\, \de^4(0)$ which we also have encountered in defining asymptotic states in quantum gravity \cite{chw4}. Noting that
\beq \label{15}
(2\pi)^4\, \La^{-4}\, \de^4(0)\sim \La^{-4}\, \intX \ar \La^{-4}\, \Vreg
\eeq
with $\Vreg$ being the regularized inner Minkowski space volume we use the fact that $\La$ is an a priori unspecified parameter which we can freely choose so that
\beq \label{16}
\La^{-4}\, \Vreg = 1.
\eeq
This is the regularization we employed in \cite{chw4} to deal with expressions of the sort of $(2\pi)^4\, \La^{-4}\, \de^4(0)$ and is the same as used in Fermi's trick to evaluate squares of Dirac's delta distribution when squaring amplitudes.

Noting that in the limit above
\beqq \label{17}
& &\quad (P_{i\al} + P_{f\al}) (P_f^\al - P_i^\al)\,
(Q_i^\be - Q_f^\be) (Q_{i\be} + Q_{f\be}) = \\
& & (P_f^2 - P_i^2)\, (Q_i^2 - Q_f^2) \ar (m^2 - m^2)\, (M^2 - M^2) = 0 \nonumber
\eeqq
vanishes we see that the inner longitudinal part of the gauge field propagator Eqn.(\ref{10}) does not contribute to the amplitude.

As there is no infrared problem for $\mu \ar 0$ we can now safely take all limits. Before doing so we also invoke the inner scale invariance of the amplitude to rescale $\La \ar \Lp$, where $\Lp = \sqrt{\Ga}$ is the Planck length in natural units $c = \hbar = 1$ and get the final expression for the scattering amplitude
\beq \label{18}
\langle p_f, q_f\, \mbox{out}\midd p_i, q_i\, \mbox{in} \rangle
= i\, (2\pi)^4\, \de^4(p_f - p_i + q_f - q_i)\,  
{\cal M}_{fi}
\eeq
with the invariant matrix element ${\cal M}_{fi}$ found to be
\beq \label{19}
{\cal M}_{fi}
= -\, (g \Lp)^2\, {\overline u}_f\, \ga^\mu\, u_i\,\,
\frac{(p_i + p_f)\cdot (q_i + q_f)}{4\,((p_f - p_i)^2 + i\ep)}\,\,
{\overline U}_f\, \ga_\mu\, U_i.
\eeq
It contains the information about the underlying dynamics of the theory and is completely symmetric under the interchange of the two particles, i.e. $p\leftrightarrow q$ and $u \leftrightarrow U$.

We note the similarity of ${\cal M}_{fi}$ with the invariant matrix element for scattering of two Dirac particles with different masses in quantum electrodynamics \cite{cli,bjd}. However, there is a crucial difference: the strength of the scattering in the case of quantum electrodynamics is proportional to the product of the two electric charges $e\, e'$ whereas in quantum gravity it is proportional to the Minkowski product of the momentum four-vectors $(g \Lp)^2\,\, \frac{(p_i + p_f)\cdot (q_i + q_f)}{4}$ which changes the dynamics completely. Note that in the rest frame of the particle with mass $M$ the coupling strenght reduces to $(g \Lp)^2\,\, \frac{(p_i + p_f)\cdot (q_i + q_f)}{4}  = (g \Lp)^2\,\, \frac{M (E + E')}{2} > (g \Lp)^2\, M\, m$.

\section{Matter-Matter Scattering Cross-Section}
In this section we calculate the cross-section for the scattering of two Dirac particles with different masses in quantum gravity to lowest order in perturbation theory.

We start with the usual Lorentz-invariant expression for the cross-section with two incoming and two outgoing Dirac fermions \cite{cli,bjd}
\beqq \label{20}
d\si &=& \frac{m\,M}{ \sqrt{(p_i\cdot q_i)^2 - m^2 M^2}}\,
\frac{m}{E_{p_f}}\frac{d^{\sl 3}p_f}{(2\pi)^3}\,
\frac{M}{E_{q_f}}\frac{d^{\sl 3}q_f}{(2\pi)^3} \\
& &\quad (2\pi)^4\, \de^4(p_f - p_i + q_f - q_i)\,
\midd {\cal M}_{fi}\midd^2. \nonumber
\eeqq

As we are interested in the unpolarized cross-section we first average over initial and sum over final states
\beq \label{21}
\midd {\cal M}_{fi}\midd^2 \ar \overline{\mid {\cal M}_{fi}\mid^2} \equiv
\frac{1}{4}\sum_{\ga_i,\ga_f;\ga'_i,\ga'_f}
\mid {\cal M}_{fi}\mid^2.
\eeq

Proceeding with the calculation of $\overline{\mid {\cal M}_{fi}\mid^2}$ we encounter two expressions of the type
\beqq \label{22}
& & \sum_{\ga_i,\ga_f} \overline u_f\,\ga^\mu\, u_i\,\,
\overline u_i\,\ga^\nu\, u_f
= \tr\, \left(\frac{p\!\!\!/_i + m}{2\, m}\, \ga^\mu\,
\frac{p\!\!\!/_f + m}{2\, m}\, \ga^\nu \right) \\
& &\quad = \frac{1}{m^2}\left(p_i^\mu p_f^\nu + p_i^\nu p_f^\mu
- p_i\cdot p_f\, \eta^{\mu\nu} + m^2\, \eta^{\mu\nu}\right). \nonumber
\eeqq
Inserting these and performing the Lorentz sums in $\overline{\mid {\cal M}_{fi}\mid^2}$ leaves us with
\beqq \label{23}
& & \overline{\mid {\cal M}_{fi}\mid^2} =
(g \Lp)^4\, \frac{\left((p_i + p_f)\cdot (q_i + q_f)\right)^2}
{16}\, \frac{1}{2\, m^2\,M^2\, \left((p_f - p_i)^2\right)^2} \\
& & \quad \left\{p_i\cdot q_i\, p_f\cdot q_f 
+ p_i\cdot q_f\, p_f\cdot q_i - m^2\, q_i\cdot q_f
- M^2\, p_i\cdot p_f +2\, m^2\, M^2 \right\}. \nonumber
\eeqq 

To further extract the physics of the two-particle scattering process we choose as coordinate system the one in which the particle with mass $M$ is at rest, i.e.
\beqq \label{24}
q_i &=& (M,0),\quad q_f = (M+E-E',\underline p - \underline p')\\
p_i &=& (E,\underline{p}),\quad
p_f = (E',\underline{p'}). \nonumber
\eeqq
Energy conservation for the chosen coordinates relates the energy $E'$ of the outgoing particle with mass $m$ to the energy $E$ of the incoming particle with mass $m$ and to the scattering angle $\th$
\beq \label{25}
E' = \frac{\frac{E}{M}  + \left(\frac{m}{M}\right)^2 + \frac{\mid\! \underline p \mid}{M} \frac{\mid\! \underline p' \mid}{M} \cos\th}{1 + \frac{E}{M}}\,M.
\eeq

Performing the phase space integrals over $q_f$ and $E'$ in Eqn.(\ref{20}) in the usual way \cite{bjd} leaves us with the scattering cross-section in the rest mass frame of the particle with mass $M$
\beq \label{26}
\frac{d\bar\si}{d\Om'} = \frac{1}{(2\pi)^2}\,
\frac{\mid \underline{p'}\mid}{\mid \underline{p}\mid}\,
\frac{m^2\, M}{M + E - \frac{\mid \underline{p}\mid}{\mid \underline{p'}\mid}\,
E'\,\cos\th}\, \overline{\mid {\cal M}_{fi}\mid^2}
\eeq
which after a little algebra can be expressed in terms of $E$ and $E'$ as
\beqq \label{27}
& &\quad \frac{d\bar\si}{d\Om'} =
\frac{(E'^2 - m^2)^{3/2}}{( E^2 - m^2)^{1/2}}\,
\frac{m^2}{E\,E' - m^2\, (1 + \frac{E}{M} - \frac{E'}{M})}\quad\quad\quad \nonumber \\
& &\quad\quad\quad\quad\quad\quad\quad \frac{(g \Lp)^4}{(4\pi)^2}\,
\frac{M^2\, (E + E')^2}{4} \\
& & \frac{1}{2\, m^2\, (E - E')^2}\,
\left\{\frac{E^2}{M^2} + \frac{E'^2}{M^2}
- \left( 1 + \frac{m^2}{M^2}\right)\,
\left(\frac{E}{M} - \frac{E'}{M}\right)\right\}. \nonumber
\eeqq
The first and last lines above are exactly the same as in the case of scattering of two Dirac particles with different masses in quantum electrodynamics \cite{bjd} whereas the middle line represents the energy-dependent gravitational interaction strength replacing the square of the fine structure constant $\al = \frac{e\, e'}{4\pi}$.

We next evaluate both the limits of a heavy scatterer $\frac{E}{M}\ll 1$ and of an ultra-relativistic incoming particle $\frac{m}{E}\ll 1$.

If the energy $E$ of the incoming particle of mass $m$ is much smaller than the mass $M$ of the scatterer Eqn.(\ref{25}) yields up to higher orders in $\frac{E}{M}$
\beq \label{28}
\frac{E}{M}\ll 1 \Rightarrow E'=E,\,
\mid\! \underline p'\mid = \mid\! \underline p\mid.
\eeq
In addition we have from Eqn.(\ref{25}) in this limit 
\beqq \label{29}
E - E' &=& 2\, \frac{\mid\! \underline p\mid^2}{M}\, \sin^2\frac{\th}{2} \\
\left\{ \dots \right\} &=& 2\, \frac{E^2}{M^2} -
2\, \frac{\mid\! \underline p\mid^2}{M^2}\, \sin^2\frac{\th}{2}, \nonumber
\eeqq
where $\left\{ \dots \right\}$ denotes the bracket appearing in Eqn.(\ref{27}).

Setting
\beq \label{30}
\be = \frac{\mid\! \underline p\mid}{E} = \mid\! \underline v\mid 
\eeq 
we find the analogue to the Mott scattering cross-section \cite{bjd} in quantum gravity expressed in terms of $\be$ and $\th$
\beq \label{31}
\frac{d\bar\si}{d\Om'} = \frac{g^4}{(4\pi)^2}\, \Ga^2\, M^2\,
\frac{ 1 - \be^2 \sin^2\frac{\th}{2}}{4\,\be^4 \sin^4\frac{\th}{2}}
\eeq
recalling that $\Lp = \sqrt{\Ga}$.

Eqn.(\ref{31}) reduces in the non-relativistic limit $\be \ar 0$ to the Rutherford-like formula
\beq \label{33}
\frac{d\bar\si}{d\Om'} = \frac{\Ga^2\, M^2\,}{4 \mid\! \underline v\mid^4 \sin^4\frac{\th}{2}}
\eeq
obtained from a quantum mechanical (and incidentially a classical) treatment of the scattering of a particle of mass $m$ off an infinitely heavy scatterer $M$ in Newtonian gravity \cite{schi} if we fix the coupling constant to be
\beq \label{32}
g^2 = 4\pi.
\eeq

Note that the value of $g^2$ depends on the conventions chosen and that it is the dimensionless combination $(g \Lp)^2\,\, \frac{M (E + E')}{2} \ll 1 $ which really matters and allows for a perturbative approach.

We again stress the fact that the scattering of two Dirac particles with different masses in the limit of a heavy scatterer and a non-relativistic incoming particle is physically equivalent to gravitational Rutherford scattering - and hence provides a non-trivial comparison and test for our claim that the theory presented in \cite{chw1,chw2,chw4} is indeed a theory describing gravity at the quantum level (allowing us in the process to fix the numerical value of $g^2$ as well).

We finally note that Eqn.(\ref{31}) does not depend on the specific properties of the incoming particle, but just on the kinematical factor $\be$ - an expression that the principle of equivalence holds in the above limit.

If on the other hand the energy $E$ is much larger than the mass $m$ of the incoming particle we have
\beq \label{34}
\frac{m}{E}\ll 1 \Rightarrow E = \mid\! \underline p\mid,\,
E' = \mid\! \underline p'\mid
\eeq
and Eqn.(\ref{25}) yields
\beqq \label{35}
E - E' &=& 2\, \frac{E\, E'}{M}\, \sin^2\frac{\th}{2} \nonumber \\
\frac{E'}{E} &=& \frac{1}{1 + 2\, \frac{E}{M}\, \sin^2\frac{\th}{2}} \\
\left\{ \dots \right\} &=& \frac{2\, E\, E'}{M^2} \left( \cos^2\frac{\th}{2}
+ \frac{2\, \frac{E^2}{M^2}\, \sin^4\frac{\th}{2}}{1 + 2\, \frac{E}{M}\, \sin^2\frac{\th}{2}} \right), \nonumber
\eeqq
where $\left\{ \dots \right\}$ again denotes the bracket appearing in Eqn.(\ref{27}).

A little algebra yields the scattering cross-section in this limit in terms of $E$ and $\th$ as
\beqq \label{36}
\frac{d\bar\si}{d\Om'} &=& \frac{(g \Lp)^4}{(4\pi)^2}\, M^2\,
\frac{1}{4\, \sin^4\frac{\th}{2}}\, 
\frac{\left( 1 + \frac{E}{M}\, \sin^2\frac{\th}{2} \right)^2}
{\left( 1 + 2\, \frac{E}{M}\, \sin^2\frac{\th}{2} \right)^3} \\
& &\quad \left( \cos^2\frac{\th}{2} + \frac{2\, \frac{E^2}{M^2}\, \sin^4\frac{\th}{2}}{1 + 2\, \frac{E}{M}\, \sin^2\frac{\th}{2}} \right).
\nonumber
\eeqq
Note that for a heavy scatterer $\frac{E}{M}\ll 1$ it reduces to Eqn.(\ref{31}) in the relativistic limit $\be \ar 1$ as it should.

We finally note that Eqn.(\ref{36}) does depend on the specific properties of the incoming particle, i.e. its mass $m$, as does the general formula Eqn.(\ref{27}) for the scattering cross-section - an expression that the principle of equivalence seems not to generally hold in a quantum context.

\section{Conclusions}
In this paper we have calculated the gravitational scattering cross-section of two Dirac particles of different masses to leading order in perturbation theory within quantum gravity described by the gauge field theory of volume-preserving diffeomorphisms. We have demonstrated that this cross-section in the limit of one very heavy particle and the other non-relativistic becomes equal to the Rutherford-like cross-section for a non-relativistic particle scattering off a Newton potential. This has allowed us to determine the value of the coupling constant appearing in the gauge field theory of volume-preserving diffeomorphisms.

This result is much less trivial than the analogous one in QED because in that case the theory describing electrodynamics at the classical and the quantum level is the same. In the case of gravity all: the invariance groups and the gauge fields, the Lagrangians, the spaces on which they are defined, the coupling mechanisms describing gravitation at the classical and the quantum level are very different - and yet the same result emerges for one of the few quantities which can be calculated in both approaches.

If indeed the gauge field theory of volume-preserving diffeomorphisms consistently describes gravity at the quantum level then this is due to the fact that the framework of relativistic quantum fields and renormalizable gauge field theories offers a very economical way to consistently implement what we know from experiment about elementary particles and their processes. It offers the necessary classical and quantum degrees of freedom to describe all: observable states labelled by a complete set of quantum numbers, causality at the micro-level, the conservation of energy, momentum and angular momentum as well as the conservation of various types of "charges" - in our case the conservation of gravitational energy-momentum and its dynamical implementation through gauging the group of volume-preserving diffeomorphisms of an inner space.

\appendix

\section{Notations and Conventions}

Generally, ({\bf M}$^{\sl 4}$,\,$\eta$) denotes the four-dimensional Minkowski space with metric $\eta=\mbox{diag}(1,-1,-1,-1)$, small letters denote spacetime coordinates and parameters and capital letters denote coordinates and parameters in inner space.

Specifically, $x^\la,y^\mu,z^\nu,\dots\,$ denote Cartesian spacetime coordinates. The small Greek indices $\la,\mu,\nu,\dots$ from the middle of the Greek alphabet run over $\sl{0,1,2,3}$. They are raised and lowered with $\eta$, i.e. $x_\mu=\eta_{\mu\nu}\, x^\nu$ etc. and transform covariantly w.r.t. the Lorentz group $SO(\sl{1,3})$. Partial differentiation w.r.t to $x^\mu$ is denoted by $\pa_\mu \equiv \frac{\pa\,\,\,}{\pa x^\mu}$.

Working in Minkowskian gauges \cite{chw2} $X^\al, Y^\be, Z^\ga,\dots\,$ denote inner Cartesian coordinates. The small Greek indices $\al,\be,\ga,\dots$ from the beginning of the Greek alphabet run again over $\sl{0,1,2,3}$. They are raised and lowered with the inner Minkowski metric $\eta$, i.e. $X_\al=\eta_{\al\be}\, X^\be$ etc. and transform covariantly w.r.t. the inner Lorentz group $SO(\sl{1,3})$. Partial differentiation w.r.t to $X^\al$ is denoted by $\nabla_\al \equiv \frac{\pa\,\,\,}{\pa X^\al}$. 

The same lower and upper indices are summed unless indicated otherwise.

All further conventions related e.g. to spinors, phase space integrals etc. are standard and taken from \cite{cli,bjd}.

\end{document}